\documentclass[epj]{svjour}
\usepackage{graphics}
\newcommand{\sitei}{\vec{i}}
\newcommand{\sitem}{\vec{m}}
\newcommand{\sitej}{\vec{j}}
\newcommand{\sitek}{\vec{k}}
\newcommand{\siteq}{\vec{q}}
\newcommand{\I}{{\rm i}}

\begin{document}
\title{Random Dispersion Approximation for the Hubbard model} 


\author{Satoshi Ejima\inst{1} 
\and 
Florian Gebhard \inst{2} 
\and 
Reinhard M.\ Noack \inst{2} 
}                     
%
%
\institute{Institute of Physics, Ernst-Moritz-Arndt Universit\"at Greifswald,
 D-17487 Greifswald, Germany
\and 
Department of Physics, Philipps-Universit\"at Marburg,
D-35032 Marburg, Germany
}
%
\date{Received: \today}
%
\abstract{We use the Random Dispersion Approximation (RDA) to study the
Mott-Hubbard transition in the Hubbard model at half band filling. 
The RDA becomes exact for the Hubbard model in infinite dimensions.
We implement the RDA on finite chains and employ the Lanczos exact 
diagonalization method in real space to calculate the ground-state energy, 
the average double occupancy, the charge gap, the momentum distribution,
and the quasi-particle weight. We find a satisfactory agreement with
perturbative results in the weak- and strong-coupling limits.
A straightforward extrapolation of the RDA data
for $L\leq 14$ lattice results in a continuous Mott-Hubbard transition at
$U_{\rm c }\approx W$. 
We discuss the significance of a possible signature
of a coexistence region between insulating and metallic
ground states in the RDA that would correspond to
the scenario of a discontinuous Mott-Hubbard transition
as found in numerical investigations 
of the Dynamical Mean-Field Theory for the Hubbard model.
\PACS{{71.10Fd}{Lattice fermion models (Hubbard model, etc.)}   
      \and
      {71.27.+a}{Strongly correlated electron systems; heavy fermions} 
      \and
      {71.30+h}{Metal-insulator transitions and other electronic transitions}
     } 
} 

\maketitle
%
\section{Introduction}
\label{intro}

The Hubbard model is the minimal lattice model
of spin-1/2 electrons that can describe the transition from a metal 
to a correlated-electron insulator. Long before the formal introduction of
the model, Mott~\cite{Mottbook} used a single-band picture 
to illustrate his idea that, at half band filling and
above some critical strength of the mutual Coulomb interaction
of the electrons, the ground state must be
an insulator with a finite gap for single-electron excitations. 

This Mott gap is readily understood
in the atomic limit where every lattice site is singly
occupied. An additional electron must create a double occupancy,
which increases the energy of the ground state by a finite amount~$U$,
the Mott gap in the atomic limit.
For itinerant electrons with bandwidth~$W$ and in 
the strong-coupling limit, $U\gg W$, 
adding a hole gives rise to the lower Hubbard band 
of width $W_{\rm hole}\approx W$ in the single-particle density of states.
Adding an electron leads to the upper Hubbard band 
of width~$W_{\rm double}\approx W$, so that the Mott gap is of the
order of $\Delta=U-W>0$. As $U/W$ becomes smaller, 
one expects a Mott transition at around $U_{\rm c}\approx W$.
Note that this is a genuine quantum phase 
transition~\cite{Gebhardbook,Sachdevbook} in the sense
that no symmetry breaking needs to occur at $U_{\rm c}$ to change the
transport properties from insulating to metallic.

These qualitative arguments do not, however, determine
the precise value of the critical interaction strength, 
the size of the gap, or the Landau
quasi-particle weight as a function of the
interaction strength. Even the nature of transition 
is not clear: are the quasi-particle weight and the gap
continuous functions of $U/W$ or do they display jump discontinuities
at the Mott-Hubbard transition? Since the transition is
fundamentally non-perturbative, 
only exact solutions can answer such questions
definitely. 

Exact solutions of interacting-electron problems
are scarce. The Hubbard model in one dimension~\cite{Fabianbook}
displays a Kosterlitz-Thouless-type transition 
because the Fermi-gas ground state 
is unstable against an infinitesimal Coulomb interaction, $U_{\rm c}=0^+$.
This nesting instability is avoided in the chiral or 
$1/r$-Hubbard model~\cite{GebhardRuckenstein}, and a continuous
transition is observed: the discontinuity
of the momentum distribution goes to zero when the gap opens
linearly above $U_{\rm c}=W$. Thus far, the Hubbard model
could not be solved exactly in any dimension $d>1$.

More insight into the Mott-Hubbard transition can be gained
in the limit of high dimensions or large lattice coordination 
number~\cite{MV,MHa}.
In this limit, the Hubbard model can be mapped onto a single-impurity
problem in a bath of fermionic particles whose properties
must be determined self-consistently 
(Dynamical Mean-Field Theory, DMFT)~\cite{Jarrell,RMP}.
The single-particle density of states for the corresponding 
single-impurity problem cannot be obtained analytically. 

An alternative is to treat the 
Hubbard model with a random dispersion relation. The
Random Dispersion Approximation (RDA) 
describes the paramagnetic phase of the
Hubbard model in infinite dimensions~\cite{Gebhardbook,RDA}.
The RDA to the Hubbard model cannot be carried out analytically either, 
i.e., numerical calculations on finite-size systems
are necessary to obtain definite results.

Unfortunately, the DMFT and the RDA approaches to the Hubbard
model in infinite dimensions give conflicting scenarios
for the Mott-Hubbard transition. All numerical studies 
of the DMFT equations favor a discontinuous transition
with a jump discontinuity in the gap, whereas the RDA results
favor a continuous transition. 

In this work, we present numerical results for the Random
Dispersion Approximation to the Hubbard model. In Section~\ref{sec:approach}
we introduce the Hubbard model, formulate the RDA approach, and give
a simple example of finite-size effects within the RDA.
In Section~\ref{sec:comparison}, we give the RDA results for
the ground-state energy,
the single-particle gap, the quasi-particle weight, and other physical
quantities in the infinite-dimensional Hubbard model.
In Section~\ref{sec:discussion}, we critically
discuss the two scenarios for the Mott-Hubbard transition
and try to reconcile the conflicting results.
A short conclusion, Section~\ref{sec:conclustions}, closes our presentation.

\section{Random Dispersion Approximation for the Hubbard model}
\label{sec:approach}

We start our presentation with the definition of the Hubbard Hamiltonian.
Next, we formulate the Random Dispersion Approximation (RDA)
for the Hubbard model
which becomes exact in the limit of infinite lattice coordination number.
Finally, we illustrate the RDA by calculating the ground-state energy
to second order in weak-coupling perturbation theory.

\subsection{Hubbard Hamiltonian}
\label{subsec:hamiltonian}

We investigate spin-1/2 electrons on a lattice. Their
motion is described by
\begin{equation}
\hat{T} =
\sum_{\sitei,\sitej;\sigma} t_{\sitei,\sitej} 
\hat{c}_{\sitei,\sigma}^{\dagger}\hat{c}_{\sitej,\sigma}^{\vphantom{\dagger}}
 \; ,
\end{equation}
where $\hat{c}^{\dagger}_{\sitei,\sigma}$,
$\hat{c}_{\sitei,\sigma}^{\vphantom{\dagger}}$ 
are creation and annihilation operators for
electrons with spin~$\sigma=\uparrow,\downarrow$ on site~$\sitei$.
Here the $t_{\sitei,\sitej}$ are the electron transfer amplitudes between
sites $\sitei$~and $\sitej$, and $t_{\sitei,\sitei}=0$.

For lattices with translational symmetry, we have $t_{\sitei,\sitej}=
t(\sitei-\sitej)$, and 
the operator for the kinetic energy is diagonal in momentum space,
\begin{eqnarray}
\hat{T} &=& \sum_{\sitek;\sigma} \epsilon(\sitek) 
\hat{c}^{\dagger}_{\sitek,\sigma}\hat{c}_{\sitek,\sigma}^{\vphantom{\dagger}}
\nonumber \; ,
\\
\epsilon(\sitek) &=& \frac{1}{L} \sum_{\sitem,\sitej} t(\sitem-\sitej)
e^{-\I (\sitem-\sitej) \sitek} \; ,
\label{disp}
\end{eqnarray}
where $\epsilon(\sitek)$ is the dispersion relation.
The electrons are assumed to interact only locally,
and the Hubbard interaction reads
\begin{equation}
\hat{D} = \sum_{\sitei} \hat{n}_{\sitei,\uparrow}\hat{n}_{\sitei,\downarrow}\; ,
\label{defD}
\end{equation}
where $\hat{n}_{\sitei,\sigma}=
\hat{c}^{\dagger}_{\sitei,\sigma}\hat{c}_{\sitei,\sigma}^{\vphantom{\dagger}}$ 
is the local density operator at site~$\sitei$ for 
spin~$\sigma$. This 
leads to the Hubbard model~\cite{HubbardI}, 
\begin{equation}
\hat{H}=\hat{T} + U \hat{D} \; .
\label{generalH}
\end{equation}
Since we are ultimately interested in the Mott-Hubbard transition,
we consider a half-filled band exclusively, i.e., 
a number of electrons~$N$ that equals the
number of lattice sites~$L$ (even). Furthermore, we consider only the
paramagnetic phase with $N_{\uparrow}=N_{\downarrow}=L/2$ electrons.

\subsection{Formulation of the approximation}
\label{subsec:Method}

\subsubsection{Hubbard model with a random dispersion relation}

The density of states for non-interacting electrons is given by
\begin{equation}
\rho(\epsilon)= \frac{1}{L} \sum_{\sitek} \delta(\epsilon-\epsilon(\sitek))
 \label{rhoepsdef}
\; .
\end{equation}
In the limit of high lattice dimensions and for translationally
invariant systems, the Hubbard model is characterized by $\rho(\epsilon)$
alone, i.e., higher-order correlation functions 
in momentum space factorize, e.g.,~\cite{PvDetal}
\begin{eqnarray}
\rho_{\siteq_1,\siteq_2}(\epsilon_1,\epsilon_2)
&\equiv&
\frac{1}{L} \sum_{\sitek} \delta\bigl(\epsilon_1-\epsilon(\sitek+\siteq_1)\bigr)
\delta\bigl(\epsilon_2-\epsilon(\sitek+\siteq_2)\bigr)
\nonumber 
\\
&=&
\rho(\epsilon_1) [ \delta_{\siteq_1,\siteq_2} 
\delta(\epsilon_1-\epsilon_2)
%
+ (1-\delta_{\siteq_1,\siteq_2}) \rho(\epsilon_2)] \, .
\nonumber \\
&&\label{RDADq} 
\end{eqnarray}
This is the characteristic property of the dispersion relation
in infinite dimensions~\cite{Gebhardbook,RDA}. 

In the Random Dispersion Approximation,
the dispersion relation $\epsilon(\sitek)$ of
the Hubbard model in~(\ref{disp}) is replaced by a random
dispersion relation $\epsilon^{\rm RDA}(\sitek)$, where
each $\sitek$~point of the Brillouin zone is randomly assigned
a kinetic energy from the probability distribution $\rho(\epsilon)$,
the bare density of states in~(\ref{rhoepsdef}).
As has been shown earlier~\cite{Gebhardbook,RDA},
the RDA for the Hubbard model is equivalent to the exact solution
of the Hubbard model in infinite lattice dimensions~\cite{MV,MHa,RMP}
when long-range order is absent.

When we choose a semi-ellipse as our bare density of states,
the RDA for the Hubbard model 
becomes exact on a Bethe lattice 
with infinite coordination number~\cite{Economou},
\begin{equation}
\rho_0(\epsilon)= \frac{4}{\pi W}\sqrt{1 -\left(\frac{2\epsilon}{W}\right)^2\,}
\quad , \quad   (|\epsilon|\leq W/2) \; ,
\label{rhozero}
\end{equation}
where $W\equiv 4t$ is the bare bandwidth. 
In the following, we shall use $t\equiv 1$
as our energy unit. 

\subsubsection{Implementation on finite systems}

In order to put the RDA into practice, 
we must calculate physical quantities for various
realizations~${\cal Q}$ of the dispersion relation
and system sizes~$L$, i.e., we must calculate expectation values
\begin{equation}
X_{\cal Q}(L)=
{}_{\cal Q}\langle \Psi_0(L) |\hat{X}|  \Psi_0(L) \rangle_{\cal Q}
\end{equation}
for observables $\hat{X}$ in the ground state $|\Psi_0(L)\rangle_{\cal Q}$
of $\hat{H}_{\cal Q}(L)$.
Physically meaningful quantities are obtained by averaging over
a large number~$R_L$ of realizations and extrapolating
to the thermodynamic limit,
\begin{equation}
X=\lim_{L\to\infty}\frac{1}{R_L}\sum_{\cal Q} X_{\cal Q}(L) \; .
\label{eq:howtoextrapolate}
\end{equation}
Typically, we choose $R_L=1000$ for
$6 \leq L\leq 14$. For distributions with a Gaussian form,
we can determine the average values with accuracy ${\cal O}(1/L)$.
The accessible system size is the limiting factor in the RDA.

We choose anti-periodic boundary 
conditions for even $L$ lattice sites in momentum space
\begin{eqnarray}
k=\frac{\pi}{L}\left(
-L-1+2m\right) \quad , \quad m=1,2,\ldots,L\; ,
\label{eqn:apbc}
\end{eqnarray}
and determine the dispersion relation $\epsilon(k)$ as the solution of
the implicit equation 
\begin{eqnarray}
 k =  \frac{4\epsilon(k)}{W} \sqrt{1-\left(\frac{2\epsilon(k)}{W}\right)^2\,}
+ 2\arcsin\left(\frac{2\epsilon(k)}{W}\right) \; .
\end{eqnarray}
This choice guarantees that we recover $\rho_0(\epsilon)$ 
from $\epsilon(k)$ in the thermodynamic limit. 
Next, we choose a permutation ${\cal Q}_{\sigma}$ for each
spin direction~$\sigma$ that permutes
the sequence $\{1,\ldots,L\}$
into $\{{\cal Q}_{\sigma}[1],\ldots,{\cal Q}_{\sigma}[L]\}$.
This defines a realization of the RDA dispersion,
${\cal Q}=[{\cal Q}_{\uparrow},{\cal Q}_{\downarrow}]$.
The numerical task is the Lanczos diagonalization of the Hamiltonian
\begin{equation}
\hat{H}_{{\cal Q}}(L)
= \sum_{\sigma}\sum_{\ell=1}^{L} \epsilon(k_{{\cal Q}_{\sigma}[\ell]})
\hat{c}_{k_{\ell},\sigma}^{\dagger}
\hat{c}_{k_{\ell},\sigma}^{\vphantom{\dagger}}
 + U \hat{D}
\; .
\end{equation}
In this way, we obtain, e.g., the momentum distribution
for a realization
\begin{equation}
n_{\cal Q}(\epsilon;L)=\frac{1}{2} \sum_{\sigma}
\left. {}_{\cal Q}\langle \Psi_0(L) | \hat{n}_{k_{\ell},\sigma} 
| \Psi_0(L) \rangle_{\cal Q} \right|_{\epsilon(k_{\ell})=\epsilon} \;,
\label{eq:nofkinRDA}
\end{equation}
which depends on momentum~$\sitek$ only via the dispersion relation
$\epsilon(\sitek)$~\cite{MV,MHa} (and depends on the realization ${\cal Q}$).

\subsection{Second-order ground-state energy in the RDA}
\label{subsec:2ndene}

\begin{figure}[ht]
\resizebox{0.95\columnwidth}{!}{%
\includegraphics{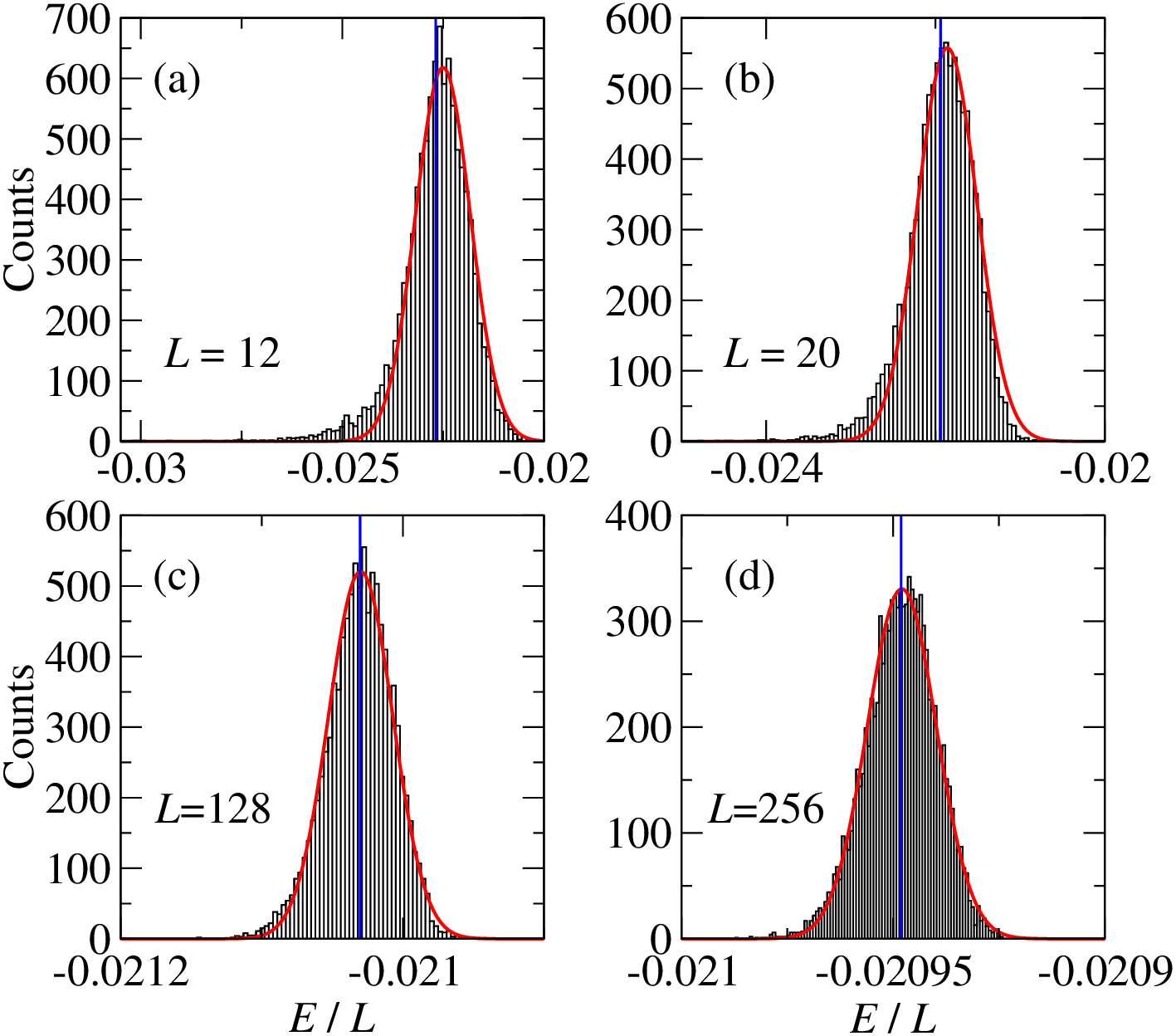}
}
\caption{Histograms of the second-order correction 
$E^{(2)}_{{\cal Q}_{\uparrow} {\cal Q}_{\downarrow}}(L)$
to the ground-state energy in weak coupling for
(a) $L=12$, (b) $L=20$, (c) $L=128$, and (d) $L=256$ sites. 
The lines give the mean values of the RDA histograms, 
and the curves are 
fits to the Gaussian function $f(E/L)=a(L) \exp[-(E/L-b(L))^2/(2c(L)^2)]$.
\label{fig:2nd-energy}}
\end{figure}

As an example, we calculate 
the ground-state energy $E=\langle \hat{H}\rangle$
to second order in the interaction strength numerically.
Within the RDA we have
(${\cal Q}=\left[ {\cal Q}_{\uparrow} {\cal Q}_{\downarrow}\right]$)
\begin{equation}
 E_{{\cal Q}_{\uparrow} {\cal Q}_{\downarrow}}(U;L)=E^{(0)}(L) +\frac{UL}{4}
  +U^2E^{(2)}_{{\cal Q}_{\uparrow} {\cal Q}_{\downarrow}}(L)+\ldots \; ,
\end{equation}
where $E^{(0)}(L)$ is the ground-state energy of the Fermi sea,
which is independent of the configuration~${\cal Q}$.
$E^{(2)}_{{\cal Q}_{\uparrow} {\cal Q}_{\downarrow}}(L)$ is the leading 
correction, which is obtained from Rayleigh-Schr\"odinger
perturbation theory as
\begin{eqnarray}
 \frac{E^{(2)}_{{\cal Q}_{\uparrow} {\cal Q}_{\downarrow}}(L)}{L}
 &=&-\left(\frac{1}{L}\right)^3
 \sum_{|k|<\pi}
 \Theta(-\epsilon_{{\cal Q}_{\uparrow}}(k))
 \sum_{|p|<\pi}
 \Theta(-\epsilon_{{\cal Q}_{\downarrow}}(p))
\nonumber \\ 
&&  \hspace*{-35pt} \sum_{0<q<2\pi}
 \frac{ \Theta(\epsilon_{{\cal Q}_{\uparrow}}(k+q))
 \Theta(\epsilon_{{\cal Q}_{\downarrow}}(p-q))
 }
 {\epsilon_{{\cal Q}_{\uparrow}}(k+q)-\epsilon_{{\cal Q}_{\uparrow}}(k)
  +\epsilon_{{\cal Q}_{\downarrow}}(p-q)-\epsilon_{{\cal Q}_{\downarrow}}(p)
 }\; ,
\nonumber \\
\label{eqn:E2}
\end{eqnarray}
where $\Theta(x)$ is the Heaviside step function.
Here $p-q$ and $k+q$ must lie in the first Brillouin zone, i.e.,
$ p-q \equiv p-q\ {\rm mod}\ 2\pi$,
$ k+q \equiv p+q\ {\rm mod}\ 2\pi$,
with
\begin{eqnarray}
 k&=&\frac{\pi}{L} \left(-L+1+2\ell\right)\quad ;
 \quad \ell=0,\ \ldots,\ L-1\; ,
 \nonumber\\
 p&=&\frac{\pi}{L} \left(-L+1+2m\right)\quad ;
 \quad m=0,\ \ldots,\ L-1\; ,
 \nonumber 
\\
 q&=&\frac{2\pi n}{L}\quad ; \quad n=1,\ \ldots,\ L-1\; .  
\end{eqnarray}
In order to illustrate the importance of finite-size effects
in the RDA, we plot histograms of the RDA data for
$E^{(2)}_{{\cal Q}_{\uparrow} {\cal Q}_{\downarrow}}(L)$ 
for $N_L=10000$ configurations and $L=12$, $20$, $128$, and $256$
in Fig.~\ref{fig:2nd-energy}. 

\begin{figure}[thbp]
\resizebox{0.95\columnwidth}{!}{\includegraphics{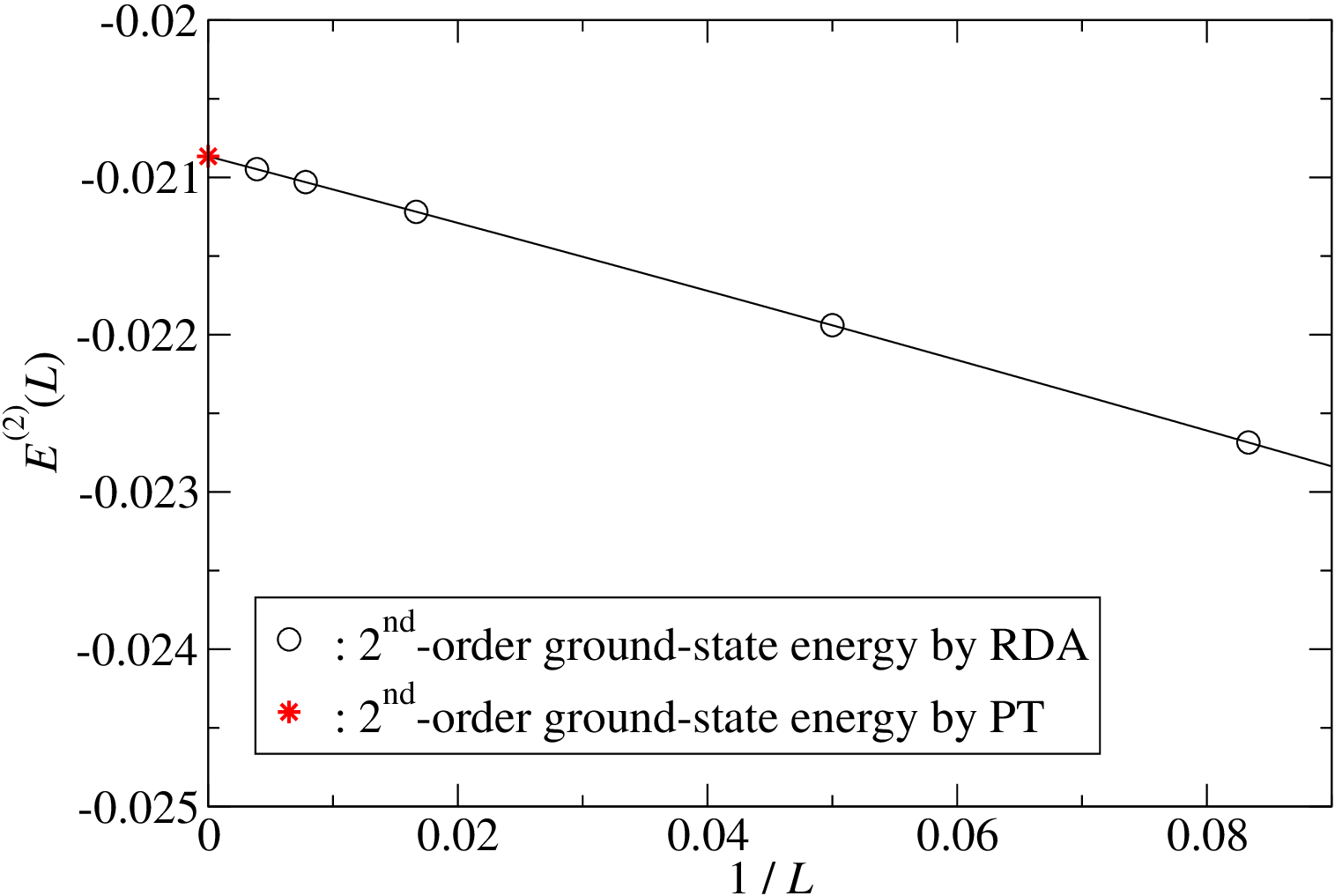}}
\caption{Finite-size extrapolation of the second-order correction
$E^{(2)}(L)$ to the ground-state energy in weak coupling.
The line is a second-order polynomial in $1/L$.
The star marks the analytic result in the thermodynamic limit,
see Eq.~(\protect\ref{eqn:ene-weak}).\label{fig:2nd-energy-extrap}}
\end{figure}

A closer analysis of Eq.~(\ref{eqn:E2}) shows that
the histograms for $E^{(2)}_{{\cal Q}_{\uparrow} {\cal Q}_{\downarrow}}(L)$
become Gaussian for large system sizes
and that their width shrinks proportionally to $1/L$. For $L<100$, 
the histograms are not perfectly Gaussian yet, introducing
a small systematic error in the mean value for
a given system size. Nevertheless, the mean value can be determined with
an accuracy that is a factor of $1/L$ better than the width of the 
distribution, $c(L)$. 
The finite-size extrapolation of the
second-order correction $E^{(2)}(L)$
is shown in Fig.~\ref{fig:2nd-energy-extrap}.

As seen from the figure, the RDA perfectly reproduces the result 
from Rayleigh-Schr\"odinger perturbation theory for the Hubbard
model in Eq.~(\ref{eqn:ene-weak}), see below. From the RDA with $L\leq 256$,
we find $E^{(2),L\leq 256}=-0.0208661(9)$, compared to the analytic result 
$E^{(2)}=-0.02086614838$. Even if we take only the results
for $L=12$ and $L=20$ into account, a fairly accurate estimate can be obtained,
$E^{(2),L\leq 20}=-0.0208(2)$.  

However, outside the perturbative limit, 
even $L=20$ is beyond our numerical capabilities.
The RDA calculations which we present in the rest of this work
are based on the exact diagonalization (ED) method which is limited
to $L\leq 14$ lattice sites because of the memory capacity. 
Calculations for $L=16$ sites using our ED code requires
about 33~GB memory. Therefore, numerical calculations for $R_L=1000$
configurations are prohibitively expensive.

In principle, larger systems can be treated using the Density-Matrix
Renormalization Group (DMRG).
However, the nonlocal nature of the hopping makes the problem
difficult in real space, and the nonlocal nature of the interaction in
momentum space makes the problem difficult in reciprocal 
space \cite{Nishimoto-kspace}.
Therefore, the maximum system size that can be treated reliably using
the DMRG is, at present, only marginally larger than the $L=14$ treated
here.

\section{Physical quantities}
\label{sec:comparison}

In this section, we present RDA data for the ground-state energy,
the average double occupancy, the charge gap, the momentum distribution,
and the quasi-particle weight. 
We compare our results with those from perturbation
theory in weak and strong coupling and from numerical approaches
to the Hubbard model in infinite dimensions (Quantum Monte Carlo,
the Dynamical DMRG, and
the Numerical Renormalization Group).

\subsection{Energy and double occupancy}
\label{subsec:ene}

The ground-state energy per site $E(U)/L$ and the average double
occupancy $d(U)$ are related by  
\begin{equation}
 d(U)=\frac{1}{L}\langle \hat{D}\rangle
     =\frac{1}{L}\frac{\partial E(U)}{\partial U}\; .
\end{equation}
In weak coupling, the ground-state energy was calculated to fourth-order 
using the Brueckner--Goldstone perturbation theory~\cite{Ruhl},
\begin{eqnarray}
\frac{E(U)}{L}&=& -\frac{8}{3\pi}+\frac{U}{4} -0.02086614838\, U^2
              		  \nonumber \\
		&&      +0.0000072475\, U^4 +{\cal O}\left(U^6\right) \; ,
\label{eqn:ene-weak}\\
  d(U)  &=& \frac{1}{4}-0.04173229676\, U
		   \nonumber \\
            &&        +0.00002899\, U^3 +{\cal O}\left(U^5\right)\; .
 \label{eqn:dU-weak}
\end{eqnarray}
In strong coupling, a diagrammatic approach 
based on the Kato--Takahashi perturbation theory provides
the results to 11th~order~\cite{EvaFlorian,NilsEva},
\begin{eqnarray}
  \frac{E(U)}{L}&=&-\frac{1}{2U}-\frac{1}{2U^3}
                -\frac{19}{8U^5}-\frac{593}{32U^7}
		\nonumber \\
           && 	-\frac{23877}{128U^9}-\frac{4496245}{2048U^{11}}
	        -{\cal O}(U^{-13})\; ,
 \label{eqn:ene-strong}\\
  d(U)  &=& \frac{1}{2U^2}+\frac{3}{2U^4}+\frac{95}{8U^6}
                +\frac{4151}{32U^8}+\frac{214893}{128U^{10}}
		\nonumber \\
	&&	+\frac{49458695}{2048U^{12}}
		+{\cal O}(U^{-14})\; .
 \label{eqn:dU-strong}
\end{eqnarray}


\begin{figure}[ht]
\resizebox{0.95\columnwidth}{!}{\includegraphics{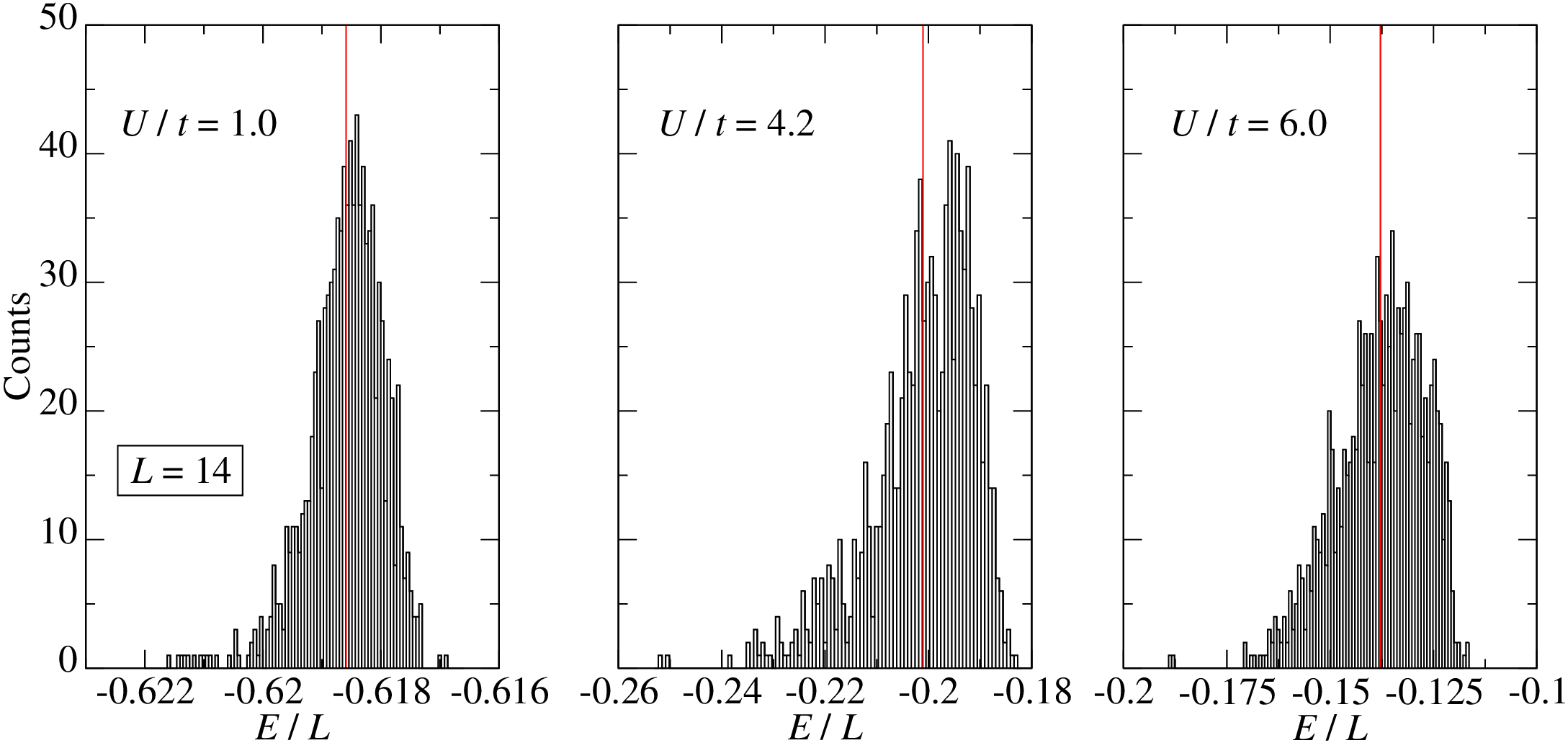}}
\caption{Histograms of the ground-state energy 
for $U/t=1$ (left), $U/t=4.2$ (middle), and $U/t=6$ (right) 
for $L=14$ (bandwidth~$W/t=4$). 
The lines indicate the mean values.\label{fig:ene-hist}}
\end{figure}

\begin{figure}[ht]
\resizebox{0.95\columnwidth}{!}{\includegraphics{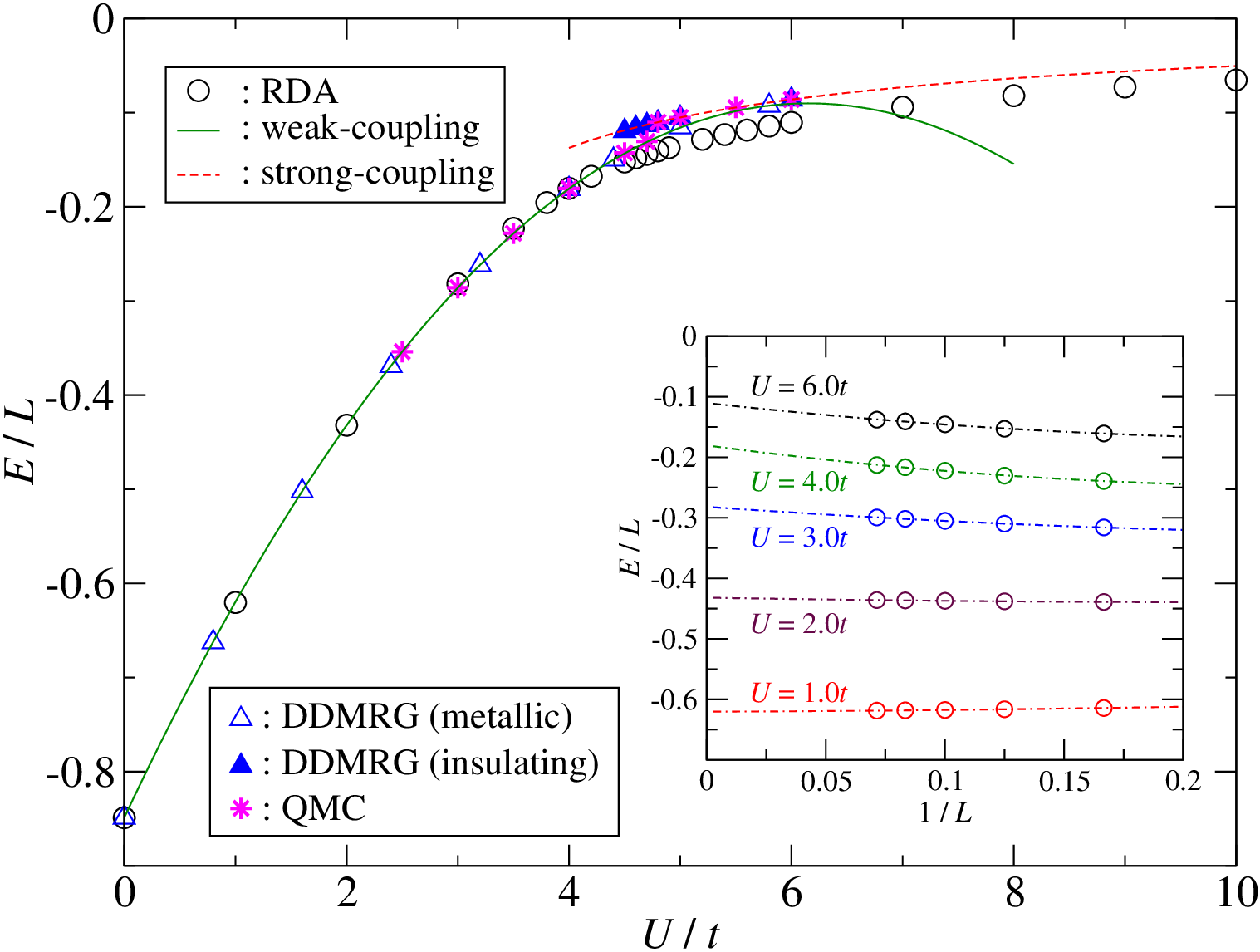}}
\caption{Extrapolated data for the ground-state energy as a
 function of the interaction strength in the RDA (circles),
 the DDMRG for a metallic solution (open triangles)~\cite{Nishim},
 the DDMRG for an insulating solution (filled triangles)~\cite{Nishim3},
 QMC (stars)~\cite{NilsEva,NilsQMC}, weak-coupling perturbation theory 
[Eq.~(\protect\ref{eqn:ene-weak})] (solid line)~\cite{Ruhl}, 
 and strong-coupling perturbation theory [Eq.\ (\protect\ref{eqn:ene-strong})]
 (dotted line) \cite{NilsEva}. 
 Inset: ground-state energy as a function of inverse system size 
 ($L\leq 14$) in the RDA for various interaction strengths.\label{fig:ene}}
\end{figure}

In Fig.~\ref{fig:ene-hist}, we show three histograms 
of the ground-state energy distribution
for $U/t=1$, $U/t=4.2$, and $U/t=6$ (bandwidth $W/t=4$) for $L=14$,
calculated using ED with the RDA. The lines indicate
the mean values. In the weak-coupling regime, the
histograms are rather symmetric 
and the relative width of the distribution is quite
small, so that the finite-size error is also small in this parameter
regime. In strong coupling, on the other
hand, the histograms are fairly asymmetric and their relative width is
much broader than in the weak-coupling region. Therefore, the
finite-size effects are much larger 
and the extrapolated data are less accurate than in the weak-coupling regime. 
The inset of Fig.~\ref{fig:ene} shows the finite-size-scaling analysis 
of the ground-state energy for various interaction strengths.
The finite-size dependence of the average values is weak,
so that reasonable estimates for the ground-state energy
can be obtained from the extrapolation of small system sizes.

In Fig.~\ref{fig:ene}, we also compare the extrapolated RDA results 
as a function of the interaction strength~$U/t$
with those from 
Quantum Monte-Carlo (QMC)~\cite{NilsEva,NilsQMC}, the Dynamical
Density-Matrix Renormalization Group method (DDMRG) \cite{Nishim,Nishim3},
weak-coupling perturbation theory,  
see Eq.\ (\ref{eqn:ene-weak}) \cite{Ruhl},
and strong-coupling perturbation theory, 
see Eq.\ (\ref{eqn:ene-strong}) \cite{NilsEva}.
As expected from the histograms and from the extrapolation of the
ground-state energy, the RDA results agree very well 
with those from all other methods in the weak-coupling regime.
However, for $U \geq W$ the agreement is worse,
as the ground-state energy extrapolated from the RDA 
is consistently lower than the ground-state energy obtained
from the DDMRG, QMC, and the strong-coupling expansion.

\begin{figure}[ht]
\resizebox{0.95\columnwidth}{!}{\includegraphics{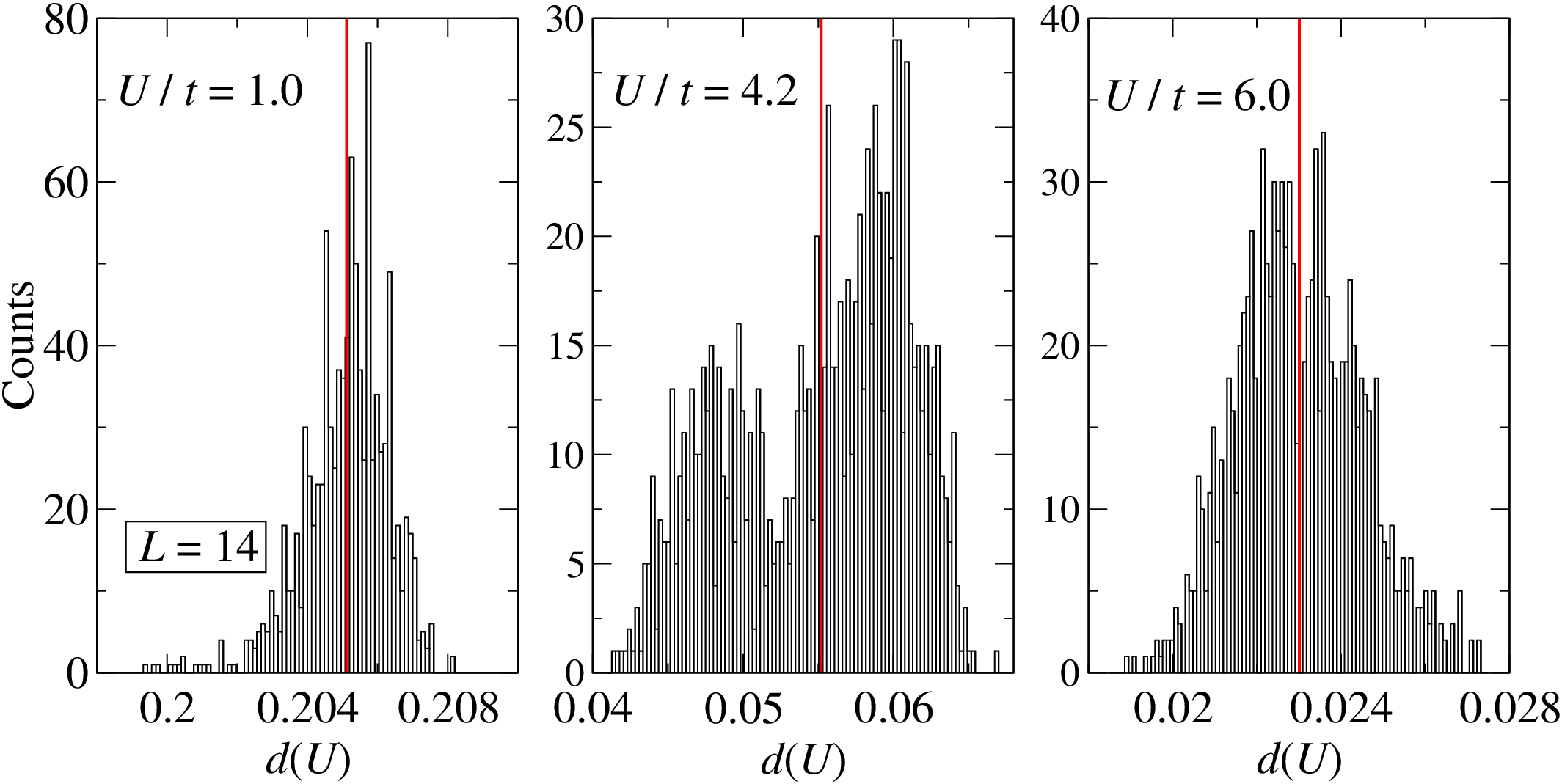}}
\caption{Histograms of the double occupancy in the RDA for 
 interaction strengths $U/t=1$ (left), $U/t=4.2$ (middle), and $U/t=6$ (right) 
for $L=14$ (bandwidth $W/t=4$). The lines indicate the mean 
values.\label{fig:dU-hist}}
\end{figure}

In Fig.~\ref{fig:dU-hist}, we show the histograms for the
double occupancy for $L=14$ sites and $U/t=1$, $U/t=4.2$, and $U/t=6$
(with bandwidth $W/t=4$). As for the ground-state energy,
the histograms are fairly narrow for small interaction strengths 
and are rather broad for large coupling. Correspondingly, the
extrapolated data are more reliable for small than for large interaction.
Interestingly, a double-peak structure appears for $U\approx W$.
This could be a signature of two coexisting solutions
for intermediate interaction strengths~\cite{RMP}.
We shall further discuss the significance of this structure in 
Section~\ref{sec:discussion}.

\begin{figure}[ht]
\resizebox{0.95\columnwidth}{!}{\includegraphics{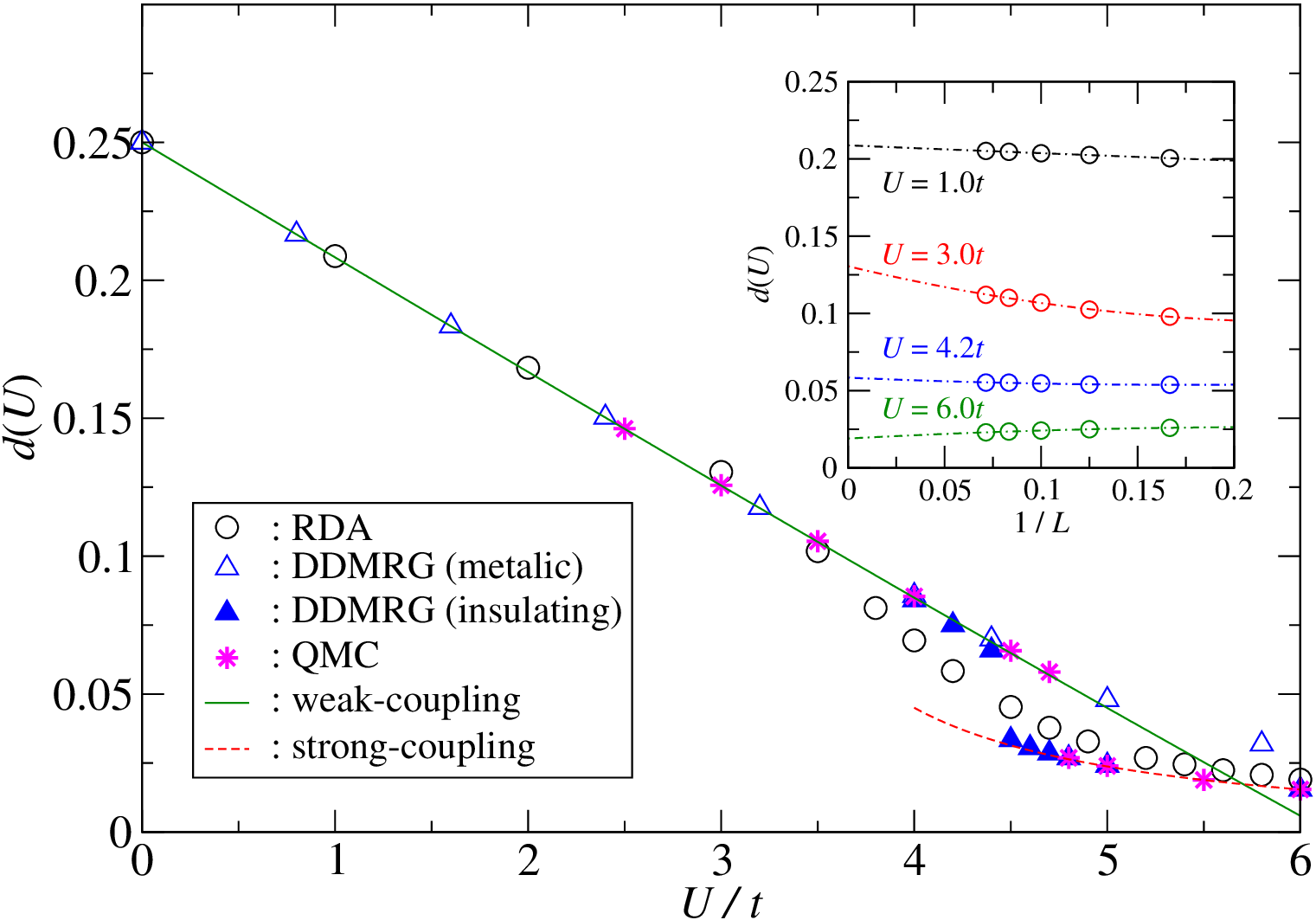}}
\caption{Extrapolated data for the double occupancy as a
 function of the interaction strength in the RDA (circles),
 the DDMRG for a metallic solution (open triangles)~\cite{Nishim},
 the DDMRG for an insulating solution (filled triangles)~\cite{Nishim3},
 QMC (stars)~\cite{NilsEva,NilsQMC}, weak-coupling perturbation theory 
 [Eq.~(\protect\ref{eqn:dU-weak})] (solid line)~\cite{Ruhl}, 
 and strong-coupling theory [Eq.~(\protect\ref{eqn:dU-strong})] 
(dotted line)~\cite{NilsEva}. 
 Inset: double occupancy as a function of the inverse system size 
 for $L\leq 14$ in the RDA for various values of the interaction 
 strengths.\label{fig:dU}}
\end{figure}

We now extrapolate the mean value of the double occupancy
as prescribed in Eq.~(\ref{eq:howtoextrapolate}).
We show the resulting data for the double occupancy in 
Fig.~\ref{fig:dU}, where we compare our results
with those from QMC~\cite{NilsEva,NilsQMC}, the DDMRG~\cite{Nishim,Nishim3},
weak-coupling perturbation theory,  
see Eq.~(\ref{eqn:dU-weak})~\cite{Ruhl},
and strong-coupling perturbation theory, 
see Eq.~(\ref{eqn:dU-strong})~\cite{NilsEva}.
Again, the RDA reproduces the weak-coupling expression very well,
and also provides a reasonable description of the strong-coupling limit.
In the transition region, $3.5 \leq U \leq 5$, the RDA result
for the double occupancy smoothly interpolates between the DDMRG and the QMC
results for the metallic phase and the insulating phase.
For a further discussion, see Section~\ref{sec:discussion}.

\subsection{Charge gap}
\label{subsec:gap}

We calculate the single-particle gap, 
\begin{equation}
 \Delta_{\cal Q}(U;L)=E_{\cal Q}(U;L+1)+E_{\cal Q}(U;L-1)-2E_{\cal Q}(U;L)
\label{eq:gapfinitesize}
\end{equation}
for each configuration ${\cal Q}$, where  $E_{\cal Q}(U;N)$ is the
ground-state energy for a system with $N$~particles. From the 
strong-coupling theory~\cite{EvaFlorian}, the charge gap 
in the thermodynamic limit is obtained as
\begin{eqnarray}
 \Delta(U)=U-4-\frac{1}{U}-\frac{3}{2U^2}
   +{\cal O}\left(\frac{1}{U^3}\right)\; .
\label{eqn:gap}
\end{eqnarray}
Again, the histograms for the charge gap are fairly broad.
We mention in passing that the ground-state momenta for different particle
numbers are not necessarily the same.

\begin{figure}[ht]
\resizebox{0.95\columnwidth}{!}{\includegraphics{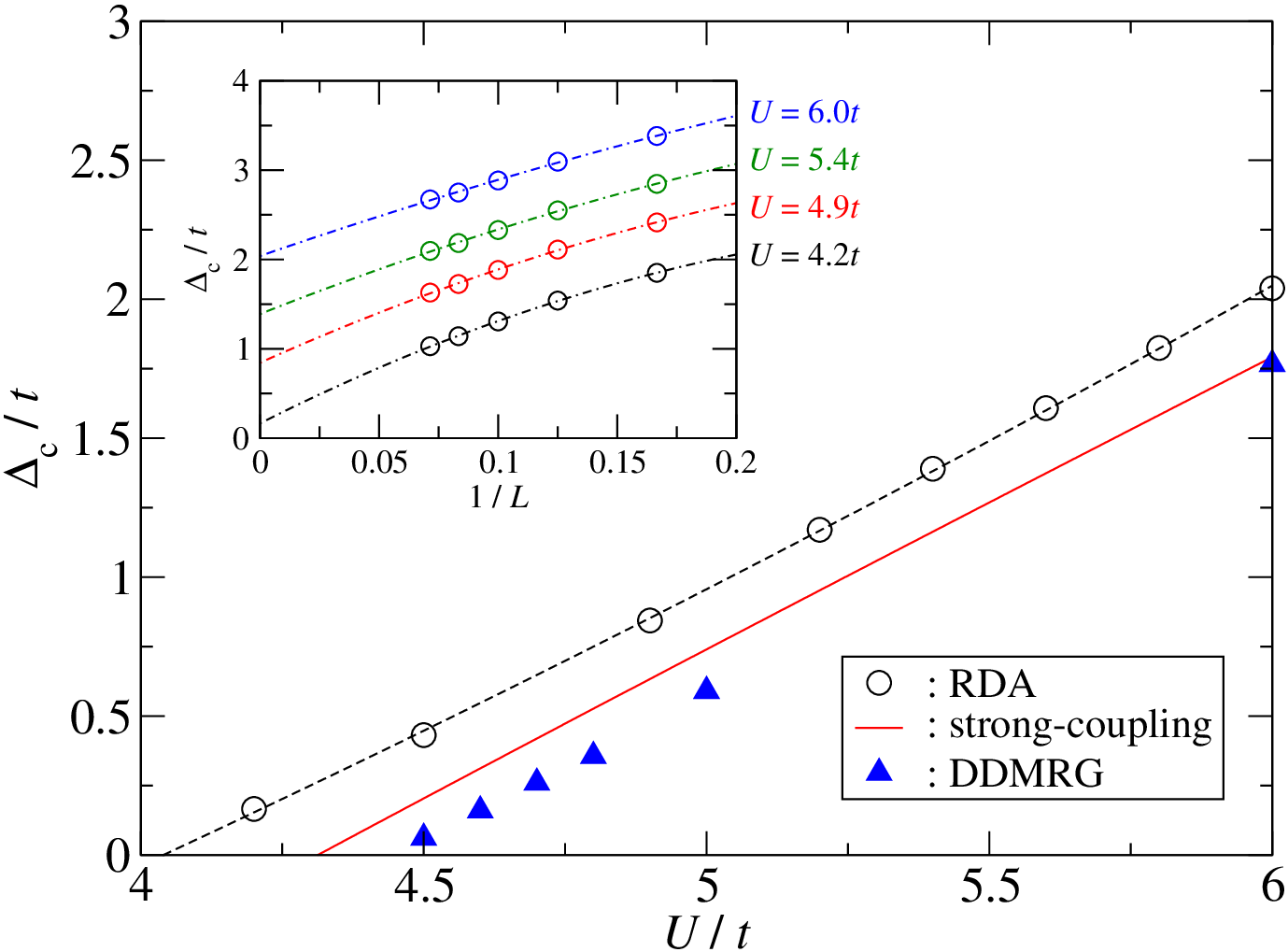}}
\caption{Extrapolated data for the single-particle gap as a
 function of the interaction strength $U$ in the RDA (circles),
the DDMRG for an insulating solution (triangles)~\cite{Nishim}, 
 and the second-order strong-coupling expansion
[Eq.~(\protect\ref{eqn:gap})] (solid line)~\cite{EvaFlorian}. 
 Inset: single-particle gap as a function of the inverse system size
 ($L\leq 14$) in the RDA for various values of the interaction 
 strengths.\label{fig:gap}}
\end{figure}

In Fig.~\ref{fig:gap}, we compare the RDA results with the
predictions from DDMRG~\cite{Nishim} and the strong-coupling expansion.
Evidently, the RDA extrapolation from data for small systems ($L\leq 14$)
overestimates the size of the gap and
underestimates the stability of the metallic solution.
Therefore, the RDA predicts a insulator-to-metal transition 
at $U\approx W$~\cite{RDA}. 

\subsection{Momentum distribution and quasi-particle weight}
\label{subsec:mom}

In the limit of infinite dimensions~\cite{MV,MHa} and within the
RDA, the momentum distribution depends on $\sitek$ 
through the bare dispersion relation $\epsilon(\sitek)$ only,
see Eq.~(\ref{eq:nofkinRDA}). 

\begin{figure}[ht]
\resizebox{0.95\columnwidth}{!}{\includegraphics{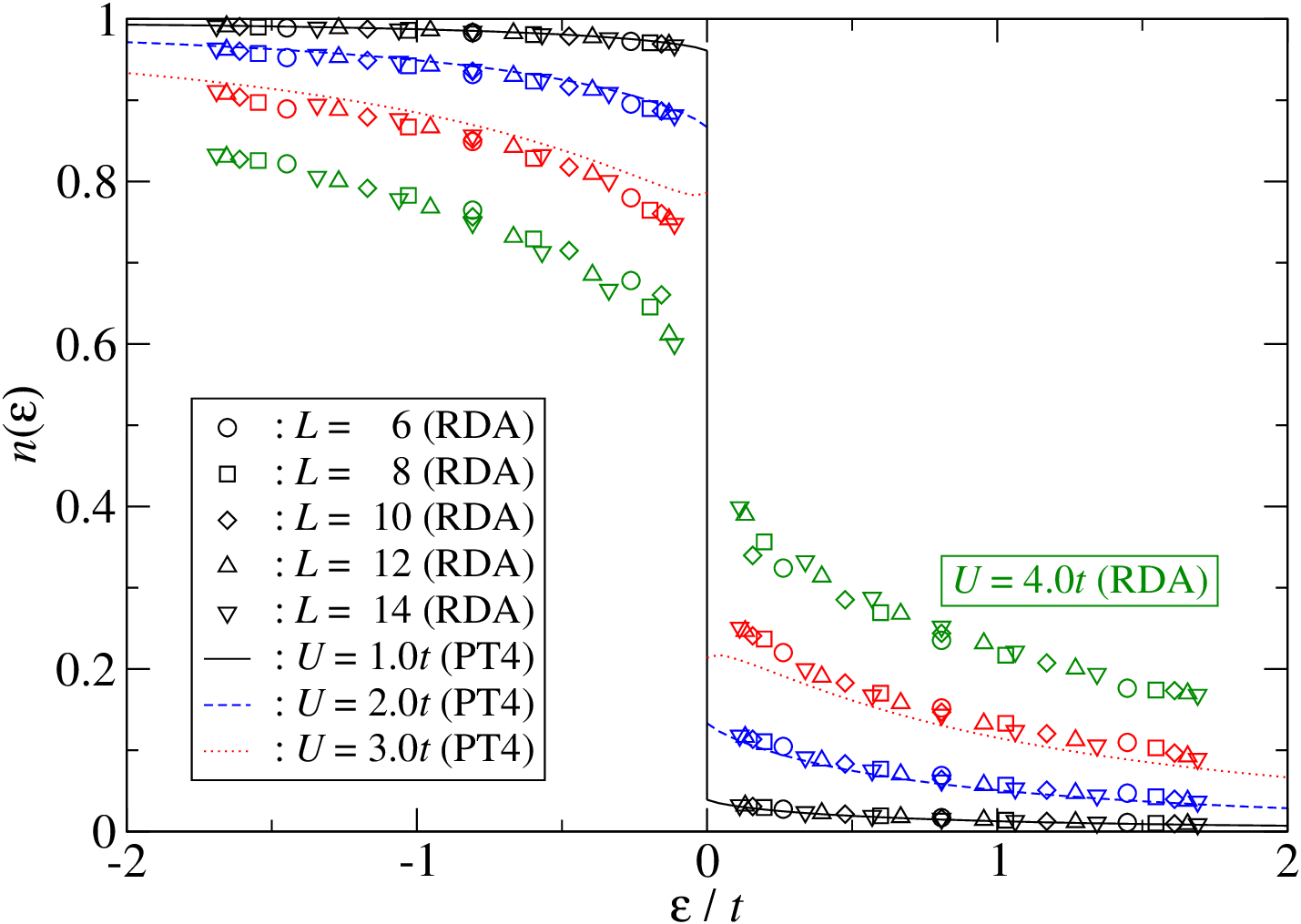}}
\caption{RDA data for the momentum distribution for various system sizes
and interaction strengths $U/t=1,2,3,4$ (symbols) in comparison
with the results from weak-coupling perturbation theory~\cite{Ruhl}
(lines).\label{fig:neps}}
\end{figure}

In Fig.~\ref{fig:neps}, we show the results for the momentum distribution
for weak to intermediate couplings. For $U\leq W/2=2$, 
we find very good agreement
between the results from the RDA and those from
weak-coupling perturbation theory
to fourth order in the interaction strength.
It is known from Feynman-Dyson perturbation theory~\cite{SandraFlorian} that
the fourth-order expression becomes unreliable for $U>0.6W=2.4$.
Therefore, we do not plot the perturbative result for $U=W$.

\begin{figure}[ht]
\resizebox{0.95\columnwidth}{!}{\includegraphics{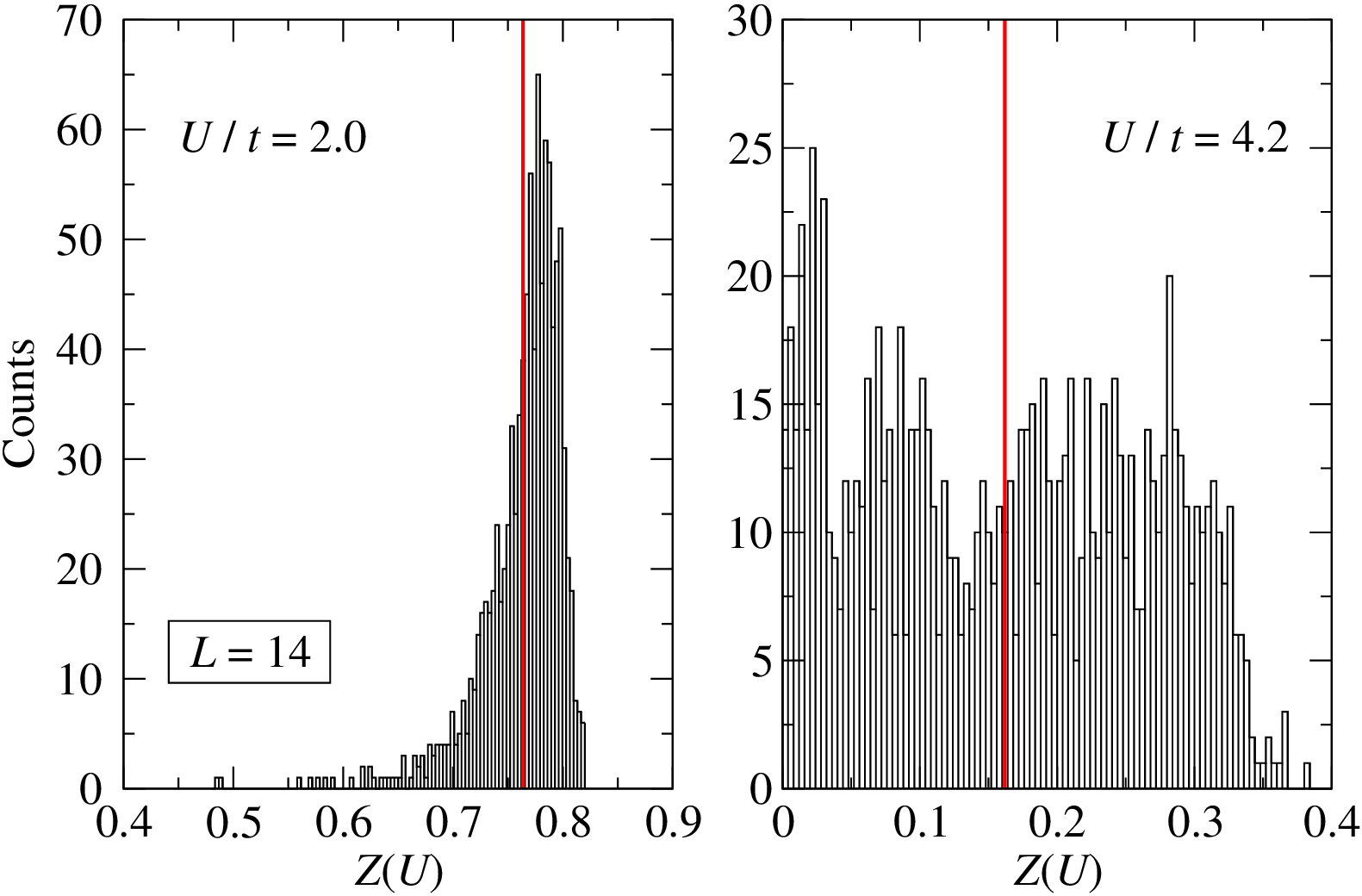}}
\caption{Histograms for the quasi-particle weight for $U/t=2$ (left)
and $U/t=4.2$ (right) for $L=14$ sites.
The long vertical lines indicate the mean values 
of the RDA histograms.\label{fig:zu-hist}}
\end{figure}

Fig.~\ref{fig:neps} shows that the finite-size effects are strongest
close to the Fermi energy, $E_{\rm F}=0$. Consequently,
a reliable calculation of
the size of the jump discontinuity at $E_{\rm F}$ is difficult.
Since the self-energy is independent
of momentum~\cite{MV,MHa,RMP},
the quasi-particle weight is obtained from
\begin{eqnarray}
 Z_{\cal Q}&=&n_{\cal Q}(\epsilon=0^-)-n_{\cal Q}(\epsilon=0^+)  \nonumber \\
&=&2 n_{\cal Q}(\epsilon=0^-) -1
\;,
\label{eqn:ZQ}
\end{eqnarray}
where the second line follows from particle-hole symmetry.
For finite system sizes, we approximate the quasi-particle weight by
the difference of the momentum distributions at $k=\pm \pi/L$ which
are closest to $\epsilon=E_{\rm F}=0$,
\begin{equation}
 Z_{\cal Q}(U;L)= 2 n_{\cal Q}\left(\epsilon(-\pi/L)\right) -1 \; ,
\end{equation}
and then extrapolate to the thermodynamic limit. 
The Brueckner-Goldstone~\cite{Ruhl}
and Feynman-Dyson~\cite{SandraFlorian} 
perturbation theories give the following result
to fourth order in the interaction strength,
\begin{equation}
 Z(U)=1-0.0817484\, U^2+0.00380158\, U^4 +{\cal O}\left(U^6\right)\; .
\label{eqn:ZU}
\end{equation}
This expression can be used for $U\leq 0.6 W=2.4$~\cite{SandraFlorian}.

In Fig.~\ref{fig:zu-hist} we show the histograms
for the quasi-particle weight for $U/t=2$ and $U/t=4.2$ for $L=14$ sites.
For small interaction strengths, the histograms show a fairly narrow peak,
and the mean value is well defined. For interaction
strengths close to the metal-insulator transition, the distribution
is broad because, for finite systems, even an `insulating' configuration 
has a finite jump in the momentum distribution, 
$Z(L;U\gg W) = {\cal O}(W/(LU))$. Therefore, 
we expect $Z_{\cal Q}(L;U/t=4.2)={\cal O}(1/L)$ which
gives $Z_{\cal Q}(L=14;U/t=4)\approx 0.07$ for our system size.
Therefore, at intermediate to large coupling strengths, the
finite-size effects for the quasi-particle  
weight are quite large.

\begin{figure}[ht]
\resizebox{0.95\columnwidth}{!}{\includegraphics{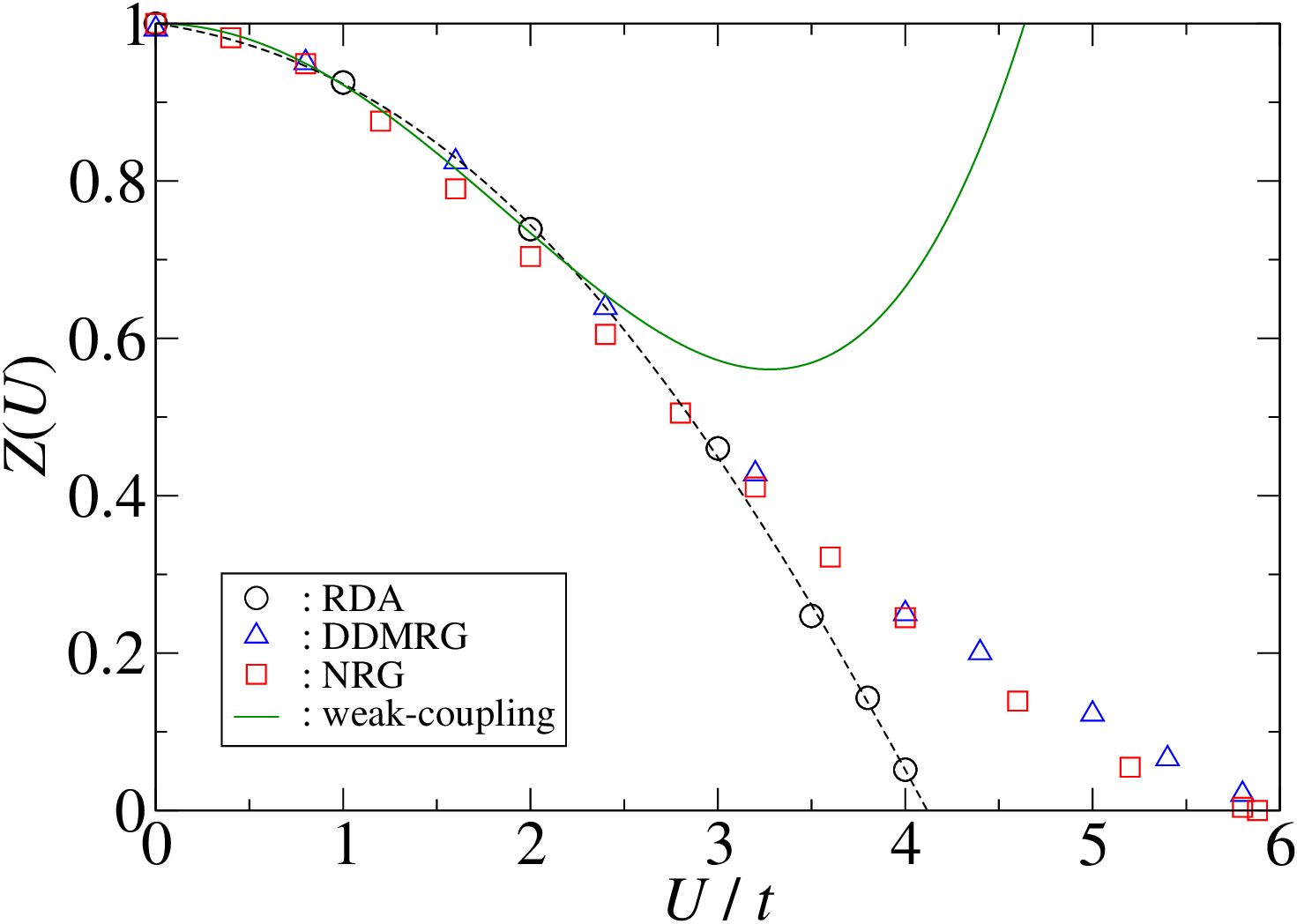}}
\caption{Extrapolated data for the quasi-particle weight as a
 function of the interaction strength in the RDA (circles), 
 the DDMRG for a metallic solution (triangles)~\cite{Nishim},
 the NRG (squares)~\cite{RBulla}, and fourth-order 
weak-coupling perturbation theory 
[Eq.~(\ref{eqn:ZU})]~\cite{Ruhl,SandraFlorian} (solid line).
 The dotted line is a guide to the eye only.\label{fig:ZU}}
\end{figure}

In Fig.~\ref{fig:ZU} we compare the extrapolated RDA results with the
predictions from DDMRG~\cite{Nishim}, NRG~\cite{RBulla}
and the weak-coupling expansion~\cite{Ruhl,SandraFlorian}.
Evidently, the RDA extrapolation for $Z(U)$
from the data for small systems ($L\leq 14$)
agrees very well with all other methods for $U\leq 3$.
For $U>3$, the RDA quasi-particle weight goes down quickly
and vanishes at $U_{\rm c}\approx W$, where the gap opens~\cite{RDA}.
All other methods used for the solution of the Dynamical Mean-Field Theory
for the Hubbard model find metallic solutions which exist up
to $U_{{\rm c},2}\approx 1.5W=6$.

We shall discuss these contrasting findings in the next section.

\section{Discussion}
\label{sec:discussion}

We first lay out the two conflicting scenarios for the Mott-Hubbard
transition in the Hubbard model with infinite coordination number.
The RDA results suggest a continuous transition, whereas
numerical treatments of the self-consistency equations
from the Dynamical Mean-Field Theory support a discontinuous transition.
We discuss objections to this scenario as well as possible
remedies. Finally, we re-analyze the RDA results 
of Section~\ref{sec:comparison} for signatures of 
coexistence of metallic and insulating ground states.

\subsection{Scenarios for the Mott transition}
\label{subsec:scenarios}

Within the RDA or in the limit of large coordination number,
the metallic phase is marked by a finite quasi-particle weight~$Z(U)$,
while a finite gap for single-particle excitations $\Delta(U)$ 
characterizes the Mott-Hubbard insulator.
Within the limits of the accuracy of the extrapolation,
the RDA predicts the Mott transition to be continuous:
as a function of the interaction strength~$U$,
the quasi-particle weight goes to zero at $U_{\rm c}\approx W$, 
where the gap opens. All thermodynamic quantities,
such as the ground-state energy 
and the average double occupancy, are continuous
functions of the interaction strength~$U$.

The scenario of a continuous transition is in conflict 
with the prediction of a discontinuous transition, as has been obtained 
from numerical solutions of the self-con\-sist\-ency equations
in the Dynamical Mean-Field Theory (DMFT)~\cite{RMP}. A metallic solution
is found for $0\leq U \leq U_{{\rm c},2}\approx 5.9\ldots 6.2$, 
whereas an insulating solution exists 
for $U_{{\rm c},1}\approx 4.5\ldots 4.8 \leq U$;
the precise values for $U_{{\rm c},1}$ and $U_{{\rm c},2}$
are still under discussion~\cite{EvaFlorian,NilsEva,RBulla,Nishim2,Uhrig}.
Since the metallic solution is found to be lower in energy than 
the insulating solution, the metal-to-insulator transition occurs
at $U_{\rm c}=U_{{\rm c},2}$ so that the quasi-particle weight goes to
zero continuously but the gap jumps to the pre-formed value associated
with the insulating solution.

A finite coexistence
region at temperature $T=0$ implies a line of first-order phase
transitions in the $(U,T)$ phase diagram 
below a critical temperature $T_{\rm c}$.
It is tempting to apply the latter to explain the phase diagram of
V$_2$O$_3$~\cite{RMP,physicstoday}, which also displays a line
of first-order phase transitions between a `metallic' 
and an `insulating' phase. 
Note, however, that the vanadium compound shows very large 
volume changes across the transition so that lattice effects
probably dominate the behavior at the transition~\cite{Krish}. 
Moreover, the critical temperature
$T_{\rm c}$ derived from the electronic mechanism alone appears to be too low
to account for the experimentally observed critical temperature.

\subsection{Problems of the scenario of a discontinuous transition}
\label{subsec:scenario-problems}

The scenario of a discontinuous transition has been criticized on 
mathematical as well as on physical grounds.

\subsubsection{Kehrein argument}

For $U\to U_{{\rm c},2}^-$,
the single-particle spectral function of the metallic solution consists
of a quasi-particle resonance of width $Z W\ll W$ 
which is energetically isolated
from the lower and upper Hubbard bands, which are split by the pre-formed 
gap~$\Delta_{\rm pre}$.
As shown by Kehrein~\cite{Kehrein}, 
a strict separation of energy scales 
implies that the self-energy
must display a singularity on an intermediate
energy scale $ZW \ll \epsilon_{\rm c}=\sqrt{Z} W\ll \Delta_{\rm pre}$.
Such a singular behavior of the self-energy is very problematic
because the DMFT mapping of the Hubbard model onto an effective
single-impurity model is based on the Dyson equation and on the skeleton
expansion around the weak-coupling limit. 
The corresponding resummations of the perturbation series
become questionable if the resulting self-energy displays
singularities as a function of frequency.

A way out of this dilemma is the concept of a quantum phase transition
`as a function of frequency'. As shown by Metzner~\cite{MetznerZPHYS},
the DMFT mapping can also be achieved perturbatively 
starting from the atomic limit. Therefore, one can argue that 
the weak-coupling perturbation theory converges for all frequencies
in the region $0 \leq U<U_{{\rm c},1}$ and that 
the strong-coupling perturbation theory converges for all frequencies 
in the region $U>U_{{\rm c},2}$. In the coexistence region,
$U_{{\rm c},1}< U < U_{{\rm c},2}$, the weak-coupling perturbation theory
converges for a finite frequency interval $|\omega|< \epsilon_{\rm c}$,
while the strong-coupling perturbation theory converges outside
this interval. Therefore, a quantum phase transition 
`as a function of frequency', signaled by a divergence in the self-energy
at $\epsilon_{\rm c}$, 
is contained in the metallic solution. It becomes a true transition
when $Z(U_{{\rm c},2})=\epsilon_{\rm c}(U_{{\rm c},2})=0$ at $U=U_{{\rm c},2}$.
Note, however, that this is a justification {\it a posteriori}, and it
is not clear {\it a priori\/} 
that numerical methods for the solution of the DMFT self-consistency
equation can treat such divergences properly.

Recently, Karski et al.~\cite{Uhrig}
proposed that bound states exist in the coexistence regime.
In their DDMRG data, they detected signatures of resonances
at the edges of the lower and upper Hubbard bands.
However, no such resonances are discernible in recent QMC data of
Bl\"umer~\cite{NilsQMC}; see, however, Ref.\ \cite{Jarrell}.
While the possibility of bound states is not discussed in Kehrein's analysis,
the existence of any kind of singularities in the density of states
makes the resummation of the perturbation series disputable.

\subsubsection{Logan-Nozi\`eres argument}

A physical argument which questions the scenario of a discontinuous transition
was presented by Logan and Nozi\-\`eres~\cite{Logan}.
In the metallic phase close to $U_{{\rm c},2}$, 
a small number of itinerant electrons, $N_{\rm itin}=ZN$,
must screen a large number $N_{\rm loc}=(1-Z)N$ of localized spins. 
For a regular Kondo screening process with Kondo energy $T_{\rm K}\approx ZW$, 
the screening energy is gained once per screening electron,
i.e., it is of the order $E_{\rm gain}=N_{\rm itin}T_{\rm K}\approx 
Z^2 W N$. The itinerant electrons must be excited over the
performed gap $\Delta_{\rm pre}$. Therefore,
the loss of energy is given by $E_{\rm loss}=
N_{\rm itin}\Delta_{\rm pre}= ZN \Delta_{\rm pre}$. 
Balancing the two energies thus gives $\Delta_{\rm pre}\approx ZW$, i.e.,
the pre-formed gap and the quasi-particle weight both vanish
at a (single) critical $U_{\rm c}$.

This argument does not take into account the possibility that
the `metallic' pre-formed gap can be larger
than the corresponding gap in the insulating solution.
Such a behavior has recently been observed in Ref.~\cite{Uhrig}.
Therefore, there is an additional energy gain 
$E_{\rm low}=ZN\Delta_{\rm pre}$ in the metallic system 
because the quasi-particle resonance pushes the filled
lower Hubbard band downwards in energy by approximately $Z\Delta_{\rm pre}$. 
This level repulsion, rather than the Kondo screening, may account 
for the existence of the quasi-particle resonance in the pre-formed gap.
Again, this is merely an {\it a posteriori\/} interpretation of the 
results obtained from numerical solutions of the self-consistency
equations in the Dynamical Mean-Field Theory.

\subsection{Coexistence of metallic and insulating ground states in the RDA?}
\label{subsec:RDA-scenario}

Kehrein's analysis~\cite{Kehrein} points to two basic problems
of the DMFT: (i) How many self-consistent solutions exist for $U>0$?
(ii) Which of the self-consistent solutions corresponds to
the Hubbard model in infinite dimensions? 

It is conceivable that multiple `metallic' and `insulating' 
solutions of the DMFT equations exist.
Some of them may easily be detected numerically,
others may be elusive. Some solutions ought to be discarded 
because their analytical properties
violate the conditions on which the mapping was based in the first place.
Therefore,
solutions of the DMFT equations that advocate a continuous Mott-Hubbard 
transition may be more difficult to find numerically~\cite{Nishim2},
but they could be relevant.

It is the main advantage of the Random Dispersion Approximation 
to the Hubbard model that
it is not based on a set of self-consistency equations.
Therefore, it provides an independent approach to the Hubbard model
with infinite coordination number.
Unfortunately, the system sizes which we can treat are fairly small,
and it is difficult to come to definite conclusions about
the existence of a continuous or a discontinuous Mott transition.

As mentioned at the end of Section~\ref{sec:comparison}, 
the RDA predicts
a continuous transition at $U_{\rm c}\approx W$~\cite{RDA}.
One may ask if and how the RDA on finite lattices
could describe a coexistence region. 
As a bare ground-state method, in the infinite-system limit
it should give a metallic solution
for $U< U_{{\rm c},2}$ and an insulating solution with a finite,
sizable gap for $U>U_{{\rm c},2}$. 
The average double occupancy would follow the weak-coupling curve
up to $U/t=5$, e.g., its value at $U/t=4.2$ would be $d(U/t=4.2)=0.077$.
When we look at the RDA histogram for the double occupancy for $L=14$ sites, 
Fig.~\ref{fig:dU-hist}, and at the corresponding finite-size
extrapolation, inset of Fig.~\ref{fig:dU}, we should
disregard the data for $d_{\cal Q}<0.053$ (configurations to the
left of the minimum 
in the histogram $d_{\cal Q}(U)$)
in order to obtain the value $d(U/t=4.2)=0.077$
from an extrapolation of the finite-size data.
The disregarded data would be the finite-size artifact of the
insulating solution with its smaller values of the average double occupancy.

\begin{figure}[ht]
\resizebox{0.95\columnwidth}{!}{\includegraphics{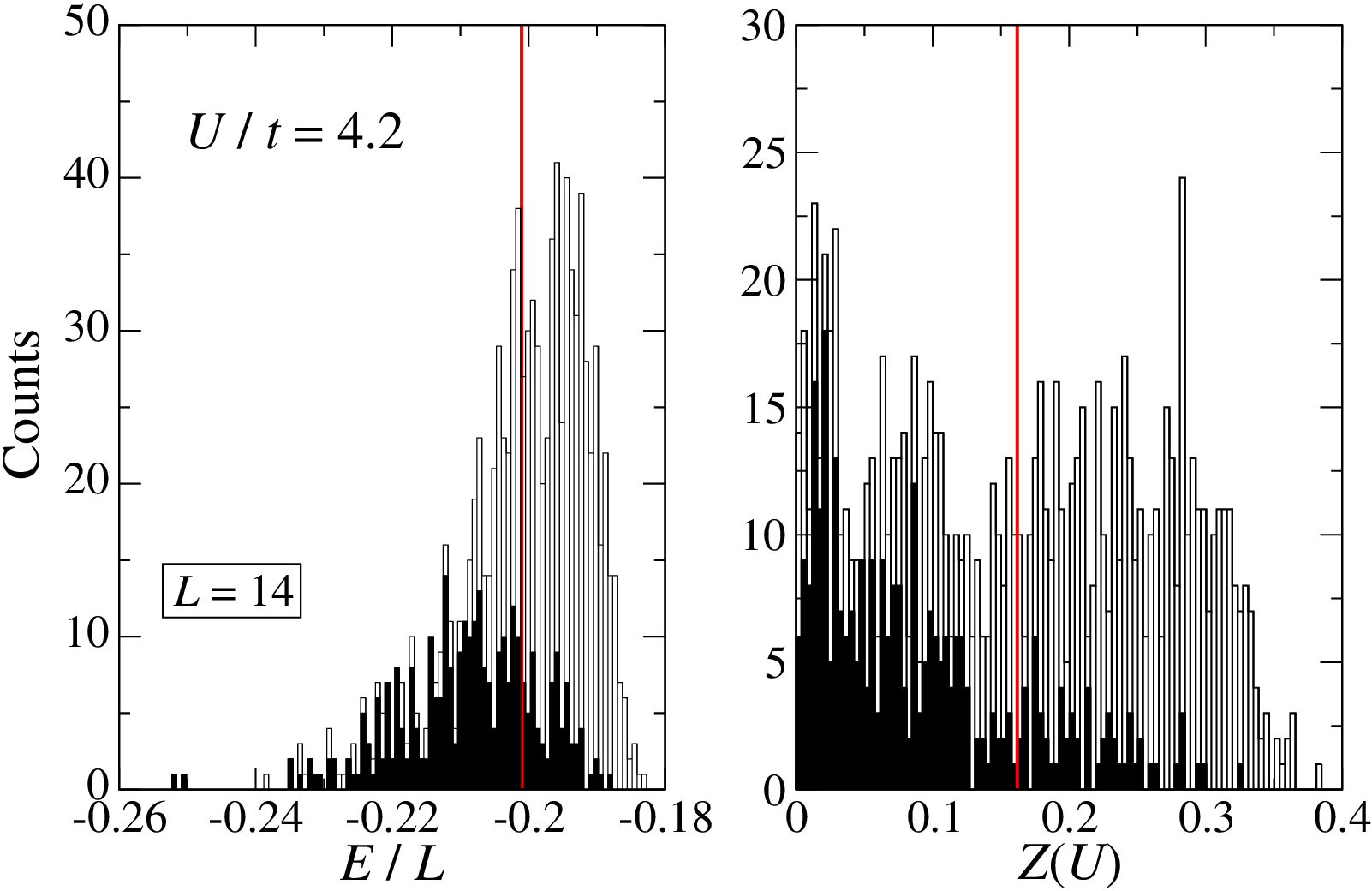}}
\caption{Histograms of the ground-state energy (left)
and the quasi-particle weight (right)
for $U/t=4.2$ and $L=14$ sites. Filled bars: configurations with
double occupancy $d_{\cal Q}<0.053$ in Fig.~\protect\ref{fig:dU-hist}; 
unfilled bars: configurations with double occupancy $d_{\cal Q}>0.053$ 
in Fig.~\protect\ref{fig:dU-hist}.\label{fig:compareZandE}}
\end{figure}

In order to test this hypothesis, we present the histograms
for the quasi-particle weight and the ground-state energy for 
$U/t=4.2$ and $L=14$ in Fig.~\ref{fig:compareZandE}.
In the figure, we distinguish between `metallic' configurations with 
$d_{\cal Q}(U/t=4.2)>0.053$ and `insulating' configurations with
$d_{\cal Q}(U/t=4.2)<0.053$. As can be seen, indeed, configurations
with a small value of $d_{\cal Q}(U/t=4.2)$ have a strong tendency to
have a small value of $Z_{\cal Q}(U/t=4.2)$, i.e., they are 
rather `insulating' than `metallic'. Correspondingly, the quasi-particle
weight will go up if we disregard the `insulating solutions'
in the average of $Z(U;L)$ and its finite-size extrapolation.

The `insulating' configurations have a \textit{lower}
ground-state energy than the `metallic' configurations.
Disregarding the `insulating' configurations would increase
the RDA prediction for the ground-state energy 
and possibly improve the agreement with the QMC data.
Note, however, that the insulating solution of the DMFT equations
has a \textit{higher} ground-state energy 
than the metallic solution in the coexistence regime. 
Therefore, the interpretation that the
disregarded configurations correspond to the insulating solution 
in the coexistence regime is problematic.

In order to investigate further the meaning of `metallic' and 
`insulating' configurations, bigger systems with more
configurations must be analyzed.
Systems with $L\leq 12$ do not show the clear 
double-peak structure for the double occupancy as seen for $L=14$
in Fig.~\ref{fig:dU-hist}. Therefore, we cannot provide a separate
finite-size extrapolation of the data for 
`metallic' and `insulating' configurations.
A `filter' which distinguishes between 
`metallic' and `insulating' configurations would introduce 
some uncontrolled bias which would act against
the construction principle on which the Random Dispersion Approximation
is based.

\section{Conclusions}
\label{sec:conclustions}

In this work, we have used the Random Dispersion Approximation (RDA) to
investigate the Hubbard model at half band filling
on a Bethe lattice with infinite coordination number.
We have employed the numerical
exact diagonalization method in real space
to calculate the ground-state energy, the average
double occupancy, the momentum distribution, and the quasi-particle weight
for 1000 realizations of the dispersion relation
on systems with $6\leq L \leq 14$
lattice sites for various interaction strengths. 
We reproduce earlier RDA results with a higher
resolution because we have used twenty times more configurations than in
previous investigations~\cite{RDA,EvaFlorian,SandraFlorian}.

In principle, the RDA describes the Hubbard model in infinite dimensions
adequately, but its usefulness is limited by the small system sizes which can
be treated using the Lanczos technique. The quantitative
agreement with results for physical quantities in the
weak-coupling and strong-coupling limits
of the Hubbard model in infinite dimensions is satisfactory,
but finite-size effects clearly reduce the significance of its
prediction of a continuous Mott-Hubbard metal-to-insulator transition
at $U_{\rm c}\approx W$ ($W$: bandwidth). 

Numerical treatments of the
Dynamical Mean-Field Theory for the Hubbard model
in infinite dimensions indicate that the Mott-Hubbard transition
is discontinuous with a finite coexistence region 
$U_{{\rm c},1}<U<U_{{\rm c},2}$ for a metallic and an insulating solution.
In the present study, we find
that the histograms, e.g., for the double occupancy, are actually
bimodal in some energy interval $W <U<1.5 W$, which could be interpreted
as the signature of a coexistence region of `insulating' and `metallic'
random configurations. In the thermodynamic limit, one
of the two peaks should shrink to zero and reveal the unique 
metallic or insulating ground state. 
In order to corroborate this hypothesis, 
much larger system sizes must be analyzed.

The RDA to the Hubbard model is free of the problems
that are generated by the mapping of the Hubbard model
in infinite dimensions to a single-impurity problem 
whose properties must be determined self-consistently.
If the RDA calculations could be done on larger system sizes,
the concept of a discontinuous Mott-Hubbard transition
could be verified independently.

\section*{Acknowledgments}
We thank Nils Bl\"umer, Satoshi Nishimoto, and Eric Jeckelmann for
useful discussions.
This work was supported in part by the Deutsche Forschungsgemeinschaft (DFG)
under grant GE 746/8-1.


\end{document}